\documentclass[conference]{IEEEtran}
\IEEEoverridecommandlockouts

\usepackage{cite}
\usepackage{amsmath,amssymb,amsfonts}
\usepackage{graphicx}
\usepackage{xcolor}
\def\BibTeX{{\rm B\kern-.05em{\sc i\kern-.025em b}\kern-.08em
    T\kern-.1667em\lower.7ex\hbox{E}\kern-.125emX}}

\usepackage{pdfpages}
\usepackage{epsfig}
\usepackage{color}
\usepackage{amsxtra}
\usepackage{enumitem}
\usepackage{acro}
\usepackage[T1]{fontenc}
\usepackage{setspace}

\usepackage[cmintegrals]{newtxmath}
\graphicspath{{Figures/}}
\usepackage{url}

\makeatletter
\def\BState{\State\hskip-\ALG@thistlm}
\makeatother

\usepackage{epstopdf}

\usepackage{amstext}

\usepackage{latexsym}
\usepackage{dsfont} 
\usepackage{color}
\usepackage{multirow}
\usepackage{float}
\usepackage[bookmarks=false]{hyperref}
\usepackage{eucal}
\usepackage{tabularx,colortbl}

\usepackage{lscape}

\usepackage{algorithm}
\usepackage{algpseudocode}

\usepackage{units}
\usepackage{cases}
\usepackage[ subrefformat=parens,labelformat=parens, caption=false, font=footnotesize]{subfig}
\usepackage{mathtools}
\usepackage{pgfplots}
\usepackage{pgf}
\usepackage{tikz}
\pgfplotsset{compat=1.15}

\newcommand{\TT}{\mathsf{T}}
\newcommand{\HH}{\mathsf{H}}

\DeclareAcronym{thz}{
  short = THz,
  long = Terahertz
}

\DeclareAcronym{ls}{
  short = LS,
  long = least square
}

\DeclareAcronym{simo}{
  short = SIMO,
  long =  single-input multiple-output
}
\DeclareAcronym{BS}{
  short = BS,
  long =  base station
}
\DeclareAcronym{UE}{
  short = UE,
  long =  user equipment
}

\DeclareAcronym{PDF}{
  short = PDF,
  long = probability density function  
}

\DeclareAcronym{ML}{
  short = ML,
  long = maximum likelihood 
}

\DeclareAcronym{ula}{
  short = ULA,
  long = uniform linear array 
}

\DeclareAcronym{los}{
  short = LoS,
  long = line-of-sight 
}

\DeclareAcronym{NMSE}{
  short = NMSE,
  long = normalized mean squared error
}

\DeclareAcronym{MSE}{
  short = MSE,
  long =  mean squared error
}

\DeclareAcronym{iid}{
  short = IID,
  long = independent and identically distributed
}

\DeclareAcronym{snr}{
  short = SNR,
  long = signal-to-noise ratio
}

\DeclareAcronym{aosa}{
  short = AoSA,
  long = array-of-subarrays
}

\DeclareAcronym{ae}{
  short = AE,
  long = array element
}

\DeclareAcronym{sa}{
  short = SA,
  long = subarray
}

\DeclareAcronym{cnn}{
  short = CNN,
  long = convolutional neural network
}

\DeclareAcronym{music}{
  short = MUSIC,
  long  = multiple signal classification
}

\DeclareAcronym{nf}{
  short = NF,
  long =  near-field
}

\DeclareAcronym{ULA}{
  short = ULA,
  long = uniform linear array 
}

\DeclareAcronym{aoa}{
  short = AoA,
  long  = angle of arrival
}

\DeclareAcronym{transmusic}{
  short = TransMUSIC,
  long  = transform multiple signal classification
}
\DeclareAcronym{esprit}{
  short = ESPRIT,
  long  = estimation of signal parameters via rotational invariance techniques
}

\DeclareAcronym{rmse}{
  short = RMSE,
  long  = root mean squared error
}

\DeclareAcronym{prmse}{
  short = PRMSE,
  long  = permutation-invariant RMSE
}

\makeatother
\makeatletter
\def\ps@IEEEtitlepagestyle{%
  \def\@oddfoot{\mycopyrightnotice}%
  \def\@oddhead{\hbox{}\@IEEEheaderstyle\leftmark\hfil\thepage}\relax
  \def\@evenhead{\@IEEEheaderstyle\thepage\hfil\leftmark\hbox{}}\relax
  \def\@evenfoot{}%
}

\def\mycopyrightnotice{%
  \begin{minipage}{\textwidth}
  \centering \scriptsize
    This work has been accepted for presentation at the IEEE SPAWC Conference. Copyright may be transferred without notice, after which this version may no longer be accessible.
  \end{minipage}
}

\begin{document}

\title{{Hierarchical THz Near-Field Localization with Subarray Processing and Covariance Correction}

{\footnotesize}
\thanks{This work was supported by the American University of Beirut (AUB) University Research Board (URB) and Vertically Integrated Projects (VIP) program and the King Abdullah University of Science and Technology (KAUST) Office of Sponsored Research (OSR) under Award No. ORFS-CRG12-2024-6478.}
}
\author{Ahmad Dkhan$^{\dagger}$, Yazan Dayoub$^{*}$, Jana El Haj$^{\dagger}$, Hadi~Sarieddeen$^{\dagger}$ \\
\small$^{\dagger}$ Department of Electrical and Computer Engineering, American University of Beirut, Beirut, Lebanon,\\
$^{*}$ Independent Researcher, Moscow, Russia\\
amd53@mail.aub.edu, dayoubyazansy@gmail.com, jje20@mail.aub.edu, hadi.sarieddeen@aub.edu.lb
}

\maketitle
\begin{abstract}
Terahertz (THz)-band near-field (NF) localization offers high spatial resolution due to short wavelengths and distance-dependent wavefront curvature in NF multi-antenna systems. However, large arrays and dense deployments, necessary to mitigate THz path loss, raise the received signal dimensionality, creating computational overhead for localization. Furthermore, traditional two-dimensional (2D) subspace algorithms suffer from excessive complexity and poor robustness under coherent sources. This paper proposes a hierarchical localization framework based on subarray (SA) processing. The first step performs 1D estimation per SA to estimate local angles. The second step combines SA outputs to estimate distances, reducing the 2D search to two 1D searches. To handle the drawback of coherent sources, a transformer-based network predicts a covariance correction, refining subspace estimation. Simulations show that the proposed hierarchical algorithm lowers complexity by four orders of magnitude. The transformer-based covariance correction improves angular accuracy by 85\% and reduces range error by $6.5~\mathrm{m}$ at $5~\mathrm{dB}$ signal-to-noise ratio in a coherent scenario compared to multiple signal classification (MUSIC).
\end{abstract}


\section{Introduction}

\ac{thz}-band communications offer ultra-high data rates and support advanced sensing and high-precision localization, which are essential for future networks and beyond~\cite{Sarieddeen2020Next,Tarboush2021Teramimo,helal2022signal}. The short wavelengths at \ac{thz} frequencies require large antenna arrays to achieve sufficient gain, resulting in highly directional beams and distance-dependent angular spread~\cite{Han2024Cross,Yang2024Performance}. Hence, an extended \ac{nf} region emerges, typically characterized by the Fraunhofer distance~\cite{Han2024Cross}, where the planar wavefront assumption no longer holds. Within this regime, the spherical wavefront introduces an additional degree of freedom, enabling distance-aware beamforming and supporting \ac{nf} localization.

Despite these advantages, \ac{nf} localization at \ac{thz} frequencies faces challenges, including severe path loss and molecular absorption~\cite{Dkhan2025THz}. While large arrays enhance signal strength and increase spatial resolution, they also significantly increase overhead and algorithmic complexity, posing challenges for real-time processing, which is critical to fully exploit the potential of \ac{thz} communications~\cite{Chen2022Tutorial}. 

A widely adopted class of algorithms for multi-source localization is based on subspace techniques, such as \ac{music}~\cite{Schmidt1986Multiple,Zhang2018Localization}, estimation of signal parameters via rotational invariance techniques (ESPRIT)~\cite{Roy2002esprit}, and Root-MUSIC~\cite{Barabell1983Improving}, which exploit the structure of the received signal covariance matrix to estimate source parameters. 

However, the performance of these algorithms depends on several assumptions, including non-coherent sources (i.e., sources whose signals are statistically uncorrelated), well-calibrated arrays, and a sufficient number of temporal snapshots~\cite{Zuo2020Subspace}. In \ac{nf} \ac{thz} arrays, these assumptions may be violated, as dense antenna spacing and signal correlation can induce source coherence and distort the array manifold. 

An alternative approach is to infer source locations directly from data using deep learning~\cite{Chen2022Tutorial}, which can adapt to complex and dynamic propagation environments where accurate channel modeling is difficult, thereby offering improved robustness. Various neural architectures have been investigated for localization, including multilayer perceptrons (MLPs)~\cite{Chen2020Deep}, \acp{cnn}~\cite{Chen2020DeepCNN}, and ResNet-based models~\cite{Cao2020Complex}. More recently, subspace-aided learning frameworks, such as TransMUSIC~\cite{ji2024transmusic} and SubspaceNet~\cite{shmuel2024subspacenet}, integrate classical subspace techniques with neural networks to enhance \ac{aoa} estimation under challenging conditions, including low-resolution measurements.

In this paper, a \ac{thz}-band hybrid-field channel model for partitioned arrays is first derived, where widely spaced subarrays are characterized by \ac{nf} spherical wavefronts, while dense subarrays follow far-field planar wavefronts. Building upon this model, a hierarchical two-step localization framework is proposed to exploit the channel structure, decomposing joint angle--range estimation into sequential one-dimensional searches and reducing computational complexity by approximately four orders of magnitude compared to 2D-MUSIC. To handle coherent sources, a transformer-based covariance correction mechanism is then developed, which predicts a full-rank surrogate covariance matrix from rank-deficient sample covariances, restoring the signal subspace and enabling robust application. Finally, simulation results validate the proposed framework, demonstrating an 85\% improvement in angular accuracy and a reduction in range error of $6.5\,\mathrm{m}$ at $5\,\mathrm{dB}$ \ac{snr} in coherent source scenarios.

Regarding notation, scalars $(a, A)$, vectors $(\mathbf{a})$, and matrices $(\mathbf{A})$ are represented by non-bold, bold lowercase, and bold uppercase letters, respectively. $\mathbf{I}_M$ is an identity matrix of size $M \times M$. $(\cdot)^\TT$, $(\cdot)^\HH$, and $(\cdot)^{-1}$ denote the transpose, Hermitian, and inverse operators, respectively. $\mathbb{E}[\cdot]$ denotes the expectation operator. $\|\cdot\|$ denotes the $\ell_2$ vector norm.

The remainder of this paper is organized as follows. Sec.~\ref{sec:system_model} describes the system model for the THz \ac{nf} partitioned array architecture. Sec.~\ref{Proposed} presents the proposed two-step localization framework, including the channel model, the generic algorithm, and the deep learning enhancement for coherent sources. Sec.~\ref{results} provides numerical results and complexity analysis. Finally, Sec.~\ref{conclusion} concludes the paper.

\section{System Model}
\label{sec:system_model}

\begin{figure}[t]
    \centering
    \includegraphics[width=0.9\linewidth]{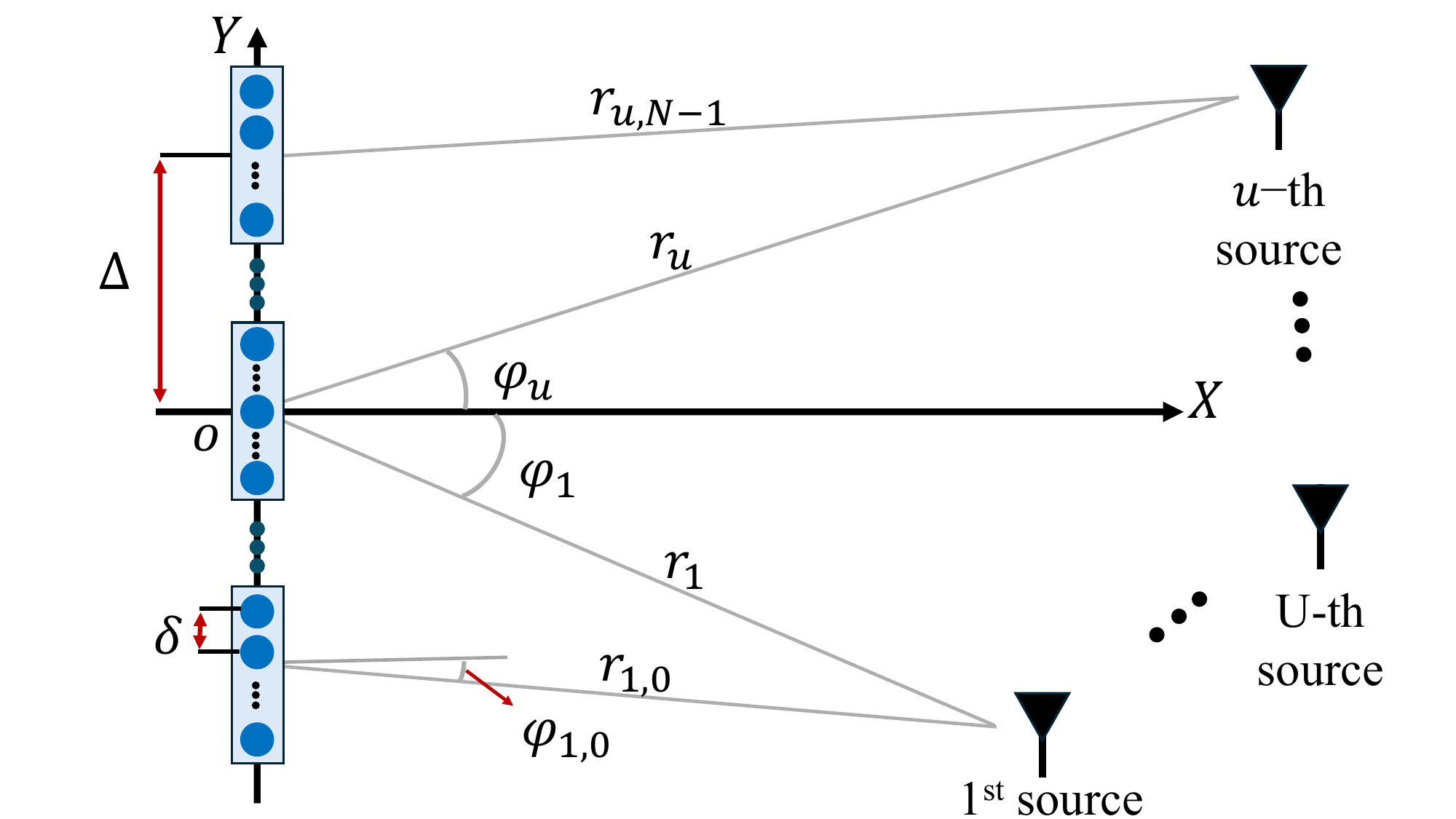}
    \caption{Layout of a partitioned ULA architecture for multi‑source localization.}
    \label{fig:sys_mod}
\end{figure}

We consider the uplink of a \ac{ula} with partitioned \acp{sa} operating at the \ac{thz} band for \ac{nf} source localization (see Fig.~\ref{fig:sys_mod}). The array comprises \(N\) \acp{sa}. The \acp{sa} are arranged along the \(y\)-axis in the Cartesian \(XY\)-plane, and the coordinate of the \(n\)th \ac{sa} is given by
\begin{equation}
    \mathbf{p}_n = [0,\, i_n \Delta]^\TT,
\end{equation}
where \(i_n = n - \frac{N - 1}{2}\) for \(n \in \{0, \dots, N-1\}\), and \(\Delta\) denotes the inter-\ac{sa} spacing. Each \ac{sa} consists of \( M \) \acp{ae}, where the coordinate of the \( m \)th element in the \( n \)th \ac{sa} is given by
\begin{equation}
    \mathbf{p}_{n,m} = [0,\, i_n \Delta + i_m \delta]^\TT,
\end{equation}
with \( i_m = m - \frac{M - 1}{2} \) for \( m \in \{0, \dots, M-1\} \), and \(\delta\) denotes the inter-element spacing within each \ac{sa}. The array aperture is \(D = (N-1)\Delta + (M-1)\delta\). The \(u\)th source is located at
\begin{equation}
\mathbf{p}_u = [r_u \cos \varphi_u, \, r_u \sin \varphi_u]^\TT, \quad u = 1,2,\dots,U,
\end{equation}
where $r_u$ is the distance from the array center and $\varphi_u$ denotes the azimuth angle measured from the $x$-axis. This work considers the sources within the radiative \ac{nf} region, enabling the joint estimation of both angle and range. The received signal at time $t = 1, 2, \dots, T$ is represented by the $MN \times 1$ vector $\mathbf{y}(t)$~\cite{Friedlander2019Localization}
\begin{equation}
\mathbf{y}(t) = \sum_{u=1}^{U} \mathbf{h}(\varphi_u, r_u) x_u(t) + \mathbf{n}(t).
\label{eq:signal_model_single}
\end{equation}
where $x_u(t)$ is the $u$th source signal, $\mathbf{n}(t) \sim \mathcal{CN}(\mathbf{0}, \sigma^2 \mathbf{I}_{MN})$ is complex Gaussian noise, and \(\mathbf{h}(\varphi,r)\!\in\mathbb{C}^{MN\times1}\) is the normalized channel vector. Stacking \(T\) snapshots, define the received signal matrix \(\mathbf{Y} = [\mathbf{y}(1), \mathbf{y}(2), \ldots, \mathbf{y}(T)] \in \mathbb{C}^{MN \times T}\), the noise matrix \(\mathbf{N} = [\mathbf{n}(1), \mathbf{n}(2), \ldots, \mathbf{n}(T)] \in \mathbb{C}^{MN \times T}\), and the source signal matrix \(\mathbf{X} = [\mathbf{x}(1), \mathbf{x}(2), \ldots, \mathbf{x}(T)] \in \mathbb{C}^{U \times T}\), where \(\mathbf{x}(t) = [x_1(t), \ldots, x_U(t)]^\top\). Then~\ref{eq:signal_model_single} can be written as
\begin{equation}   
\mathbf{Y} = \mathbf{H}(\boldsymbol{\varphi}, \mathbf{r}) \mathbf{X} + \mathbf{N},
\label{eq:signal_model}
\end{equation}
with \(\boldsymbol{\varphi} = [\varphi_1,\ldots,\varphi_U]^\top\), \(\mathbf{r} = [r_1,\ldots,r_U]^\top\), and the channel matrix \(\mathbf{H}(\boldsymbol{\varphi},\mathbf{r}) = [\mathbf{h}(\varphi_1,r_1), \ldots, \mathbf{h}(\varphi_U,r_U)] \in \mathbb{C}^{MN \times U}\).

The localization problem can be formulated as estimating the angles and ranges of the \(U\) sources from \(\mathbf{Y}\). The traditional 2D-\ac{music} approach for \ac{nf} localization jointly estimates both angle and range over the entire array\cite{Schmidt1986Multiple}. While this method can achieve high resolution, it struggles with extremely high computational complexity, especially for large arrays and fine angle-range grids. Specifically, for a base station with $MN$ antennas, $T$ snapshots, an angle grid of $N_\varphi$ points, and a range grid of $N_r$ points, the total complexity is given by
\(\mathcal{O}\Big((MN)^3 + T (MN)^2 + N_\varphi N_r (MN)^2 \Big)\). This high complexity motivates alternative approaches that reduce computation while preserving accurate \ac{nf} localization. 

\section{Proposed Localization Framework}
\label{Proposed}

\subsection{Channel model}
\label{subsec:channel_model}
We adopt the intra-subarray-far-field channel model, assuming the source is far-field relative to each \ac{sa}\footnote{Within each \ac{sa}, the element spacing $\delta$ is small compared to the local \ac{sa} range $r_{u,n}$, i.e., \(\frac{|i_m|\delta}{r_{u,n}} \ll 1.\)}, while inter-\ac{sa} propagation occurs in the \ac{nf}. The distance between the $n$th \ac{sa} and the $u$th source is
\begin{equation}
\begin{aligned}
r_{u,n} &= \|\mathbf{p}_{n} - \mathbf{p}_u\|
= \sqrt{r_u^2 + i_n^2 \Delta^2 - 2 r_u i_n \Delta \sin\varphi_u} \\
&= r_u \sqrt{1 + \left(\frac{i_n \Delta}{r_u}\right)^2 - 2 \frac{i_n \Delta}{r_u} \sin\varphi_u}.
\end{aligned}
\end{equation}
By applying a second-order Taylor approximation, $\sqrt{1+\epsilon} \approx 1 + \frac{\epsilon}{2} - \frac{\epsilon^2}{8}$ for small $\epsilon$, $r_{u,n}$ can be approximated as ~\cite{Zhang2018Localization}
\begin{equation}
\label{eq:sec_taylor}
r_{u,n} \approx   
r_u - i_n \Delta\sin\varphi_u 
+ \frac{i_n^2 \Delta^2}{2r_u} \cos^2\varphi_u.
\end{equation}
The local incidence angle $\varphi_{u,n}$ at \ac{sa} $n$ is defined as

\begin{equation}
\sin\varphi_{u,n} = \frac{r_u \sin\varphi_u - i_n \Delta}{r_{u,n}}.
\end{equation}
The distance between the $m$th \ac{ae} in the $n$th \ac{sa} and the $u$th source is approximated using a first-order Taylor expansion as

\begin{equation}
\begin{aligned}
r_{u,n}^{(m)} &= \sqrt{\,r_{u,n}^2 + (i_m\delta)^2 - 2\,i_m\delta\,r_{u,n}\sin\varphi_{u,n}\,} \\[2pt]
&\approx r_{u,n} - i_m\delta\sin\varphi_{u,n}.
\end{aligned}
\end{equation}
The channel vector for the \(n\)th \ac{sa} is~\cite{Tarboush2021Teramimo}

\begin{equation}
[\mathbf{h}(\varphi_u,r_u)]_n =
\frac{r_u}{r_{u,n}} 
\, e^{-\mathcal{K}(f_c)\,(r_{u,n}-r_u)}
\, e^{-j \frac{2\pi}{\lambda} (r_{u,n}-r_u)}
\, \mathbf{a}(\varphi_{u,n}),
\label{eq:channel_model_vector_clean}
\end{equation}
where $\mathcal{K}(f_c)$ is the molecular absorption coefficient\footnote{Absorption mainly results from water vapor resonance lines in the \ac{thz} band, causing frequency-dependent path loss (details in~\cite{Tarboush2021Teramimo}).}, and $\mathbf{a}(\varphi_{u,n})$ denotes the array response vector of the $n$th \ac{sa}, modeling the relative phase shifts induced by a local planar wave across its \ac{ae} elements:
 \begin{equation}
\mathbf{a}(\varphi_{u,n}) =
\left[
e^{j \frac{2\pi}{\lambda} i_0 \delta \sin\varphi_{u,n}}, \;\;
\cdots, \;\;
e^{j \frac{2\pi}{\lambda} i_{M-1} \delta \sin\varphi_{u,n}}
\right]^\TT \in \mathbb{C}^{M \times 1}.
\end{equation}

\subsection{Proposed Localization Framework}

We propose a hierarchical \ac{nf} localization framework for multi-user systems, summarized in Algorithm~\ref{alg:multiuser_two_step}. Each \ac{sa} is processed independently, with received signal matrix $\mathbf{Y}_n$. The local covariance matrix is computed as
\begin{equation}
    \mathbf{R}_n = \mathbb{E}\left[ \mathbf{y}_n \mathbf{y}_n^\mathsf{H} \right] = \frac{1}{T} \mathbf{Y}_n \mathbf{Y}_n^\HH,
\end{equation}
with complexity $\mathcal{O}(N T M^2)$. Local angles $\{\hat{\varphi}_{n,u}\}_{u=1}^U$ are estimated using a 1D method (e.g., MUSIC, Root‑MUSIC, etc.); the complexity for MUSIC is $\mathcal{O}(N (M^3 + N_\varphi M^2))$. The estimated angles are then used to combine the received signals toward each source direction,
\begin{equation}
    \mathbf{y}_{n,u} = \mathbf{a}(\hat{\varphi}_{n,u})^\HH \mathbf{Y}_n,
\end{equation}
with complexity $\mathcal{O}(N U M T)$. The combined signals across all \acp{sa} for each user are
\begin{equation}
    \mathbf{Y}_{u,\mathrm{comb}} = [\mathbf{y}_{1,u}^\TT, \dots, \mathbf{y}_{N,u}^\TT]^\TT,
\end{equation}
and global \acp{aoa} are taken from a designated reference \ac{sa} (typically the center \ac{sa}). For each source $u$, the covariance of the combined signals,
\begin{equation}
    \mathbf{R}_u = \frac{1}{T} \mathbf{Y}_{u,\mathrm{comb}} \mathbf{Y}_{u,\mathrm{comb}}^\HH,
\end{equation}
is computed (complexity $\mathcal{O}(U T N^2)$), and the range $\hat{r}_u$ is estimated with the reference \ac{aoa} $\hat{\varphi}_u$ (complexity $\mathcal{O}(U (N^3 + N_r N^2))$ for \ac{music}). This hierarchical procedure decouples angle and range estimation, allowing independent SA processing while maintaining global consistency with total complexity \(  
\mathcal{O}\Big(N(T M^2 + M^3 + N_\varphi M^2) + N U M T + U(T N^2 + N^3 + N_r N^2)\Big).
\)
Moreover, combining \ac{sa} outputs along estimated directions improves effective \ac{snr}, enabling accurate multi-user \ac{nf} localization.

\begin{algorithm}[ht]
\caption{Generic Multi-User Hierarchical NF Localization Framework with \ac{sa} Processing}
\label{alg:multiuser_two_step}
\begin{algorithmic}[1]
\Require Array parameters $\{N, M, \lambda, \delta, \Delta\}$, number of sources $U$, and received data $\mathbf{Y}$
\Ensure Estimated positions $\{(\hat{r}_u, \hat{\varphi}_u)\}_{u=1}^U$

\State \textbf{Step 1: Local Angle Estimation}
\For{each \ac{sa} $n=1,\dots,N$}
    \State Compute local covariance $\mathbf{R}_n = \tfrac{1}{T}\mathbf{Y}_n\mathbf{Y}_n^\HH$
    \State Estimate local angles $\{\hat{\varphi}_{n,u}\}_{u=1}^U$
\EndFor

\State \textbf{Step 2: Range Estimation}
\For{each source $u=1,\dots,U$}
    \State Select reference \ac{sa} $n_\mathrm{ref}$ and set $\hat{\varphi}_u = \hat{\varphi}_{n_\mathrm{ref},u}$
    \State Form \ac{sa} outputs $\mathbf{y}_{n,u} = \mathbf{a}(\hat{\varphi}_{n,u})^\HH\mathbf{Y}_n$
    \State Combine them as $\mathbf{Y}_{u,\mathrm{comb}} = [\mathbf{y}_{1,u}^\TT,\dots,\mathbf{y}_{N,u}^\TT]^\TT\!\in\!\mathbb{C}^{N\times T}$
    \State Compute covariance $\mathbf{R}_{u} = \tfrac{1}{T}\mathbf{Y}_{u,\mathrm{comb}}\mathbf{Y}_{u,\mathrm{comb}}^\HH$
    \State Estimate range $\hat{r}_u$ using $\mathbf{R}_{u}$ and $\hat{\varphi}_u$
\EndFor

\end{algorithmic}
\end{algorithm}

\subsection{Proposed Deep Learning Enhancement}

While the hierarchical framework significantly reduces computational complexity, its performance deteriorates in the presence of coherent sources due to subspace collapse. To address this limitation, we introduce a learning-based covariance correction mechanism.

Subspace-based methods such as \ac{music} and Root-MUSIC rely on the assumption of uncorrelated sources, under which the source covariance $\mathbf{R}_n$ is diagonal and full-rank, enabling reliable separation of signal and noise subspaces. However, for coherent sources, $\mathbf{R}_n$ becomes non-diagonal and rank-deficient, leading to a breakdown in eigen-decomposition and degraded \ac{aoa} estimation performance. Classical decorrelation techniques, such as spatial smoothing, partially mitigate this issue but at the cost of reduced effective array aperture and limited resolvability of coherent sources~\cite{Zhao2024Augmented}.

\begin{figure}[t]
    \centering
    \includegraphics[width=0.95\linewidth]{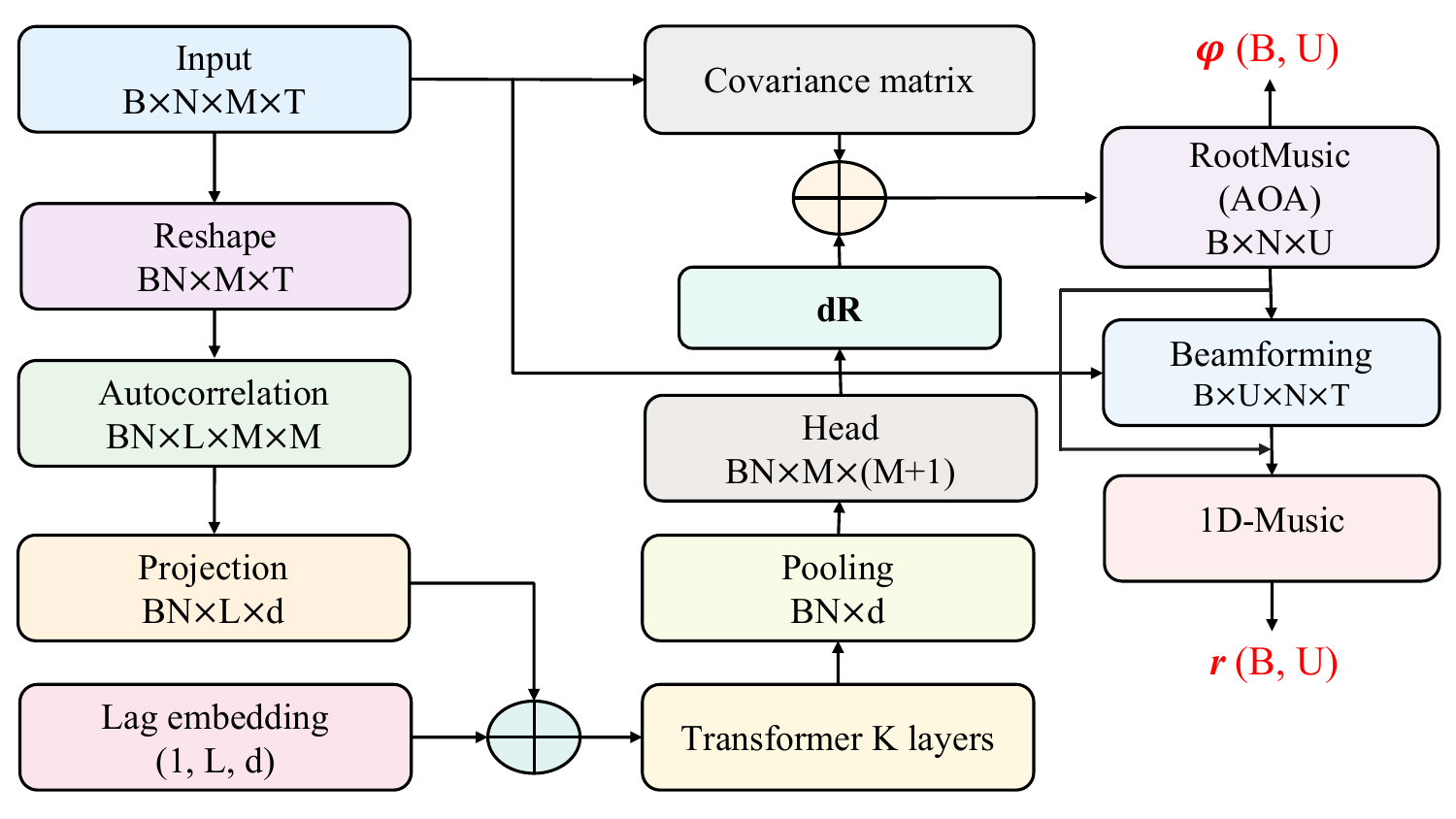}
    \caption{Overview of the proposed AI-aided sources' location estimation pipeline.}
    \label{fig:model_architecture}
\end{figure}

To overcome these limitations, we adopt a transformer-based approach that learns to correct the sample covariance matrix before subspace processing. As illustrated in Fig.~\ref{fig:model_architecture}, the network processes multi-lag covariance representations and outputs a correction matrix that is added to the sample covariance prior to subspace decomposition. The proposed model operates on each \ac{sa} independently and leverages temporal structure captured through multiple autocorrelation matrices computed over a set of time lags $\tau \in \mathcal{T}$. Specifically, the lagged autocorrelation matrices are defined as

\begin{equation}
    \mathbf{R}_{n,\tau} = \frac{1}{T - \tau} \sum_{t=\tau}^{T-1} \mathbf{Y}_{n}[:, t] \, \mathbf{Y}_{n}[:, t-\tau]^\HH.
\end{equation}
The real and imaginary parts of each $\mathbf{R}_{n,\tau}$ are concatenated along the channel dimension to form a real-valued input tensor of size $M \times M \times (2|\mathcal{T}|)$. This tensor is then flattened, projected into a \(d\)-dimensional feature space, and augmented with learnable time-lag embeddings before being processed by a transformer encoder. The transformer captures inter-lag relationships and temporal dependencies. The resulting token representations are averaged. Let $\mathbf{dR} \in \mathbb{C}^{M \times M}$ be the covariance correction matrix. The model directly predicts $\mathbf{dR}$ (via its Cholesky decomposition to ensure positive definiteness), rather than regressing the full surrogate covariance matrix, thereby simplifying the learning task.

The mapping from input covariance to the reconstructed covariance matrix is defined as
\begin{equation}
    \hat{\mathbf{R}}_{n} = \mathbf{R}_{n,0} + f_{\theta}\bigl(\{\mathbf{R}_{n,\tau} : \tau \in \mathcal{T}\}\bigr),
\end{equation}
where $f_{\theta}$ denotes the shared neural network parameterized by $\theta$. 
The justification behind this design is twofold. First, the transformer learns to reconstruct a full‑rank covariance matrix from the rank‑deficient sample covariance of coherent sources, thereby restoring the correct signal subspace dimension. Second, the predicted correction $\mathbf{dR}$ compensates for the distortion caused by correlated source signals, effectively decorrelating the sources before subspace decomposition. This allows Root‑MUSIC to separate signal and noise subspaces accurately, whereas classical methods fail due to subspace collapse.
Each reconstructed covariance matrix $\hat{\mathbf{R}}_{n}$ is then independently passed to the Root-MUSIC algorithm to produce the local \acp{aoa} estimates. Root-MUSIC identifies candidate \ac{aoa} as the angles $\varphi$ satisfying
\begin{equation}
    \mathbf{a}^\HH(\varphi) \, \mathbf{E}_n \, \mathbf{E}_n^\HH \, \mathbf{a}(\varphi) = 0,
\end{equation}
where the $\mathbf{a}(\varphi)$ surrogate is the $M$-element steering vector and $\mathbf{E}_n$ is the noise subspace from the eigendecomposition of $\hat{\mathbf{R}}_{n}$. During training, the model is optimized end-to-end with a permutation-invariant periodic \ac{rmse} loss over the \ac{aoa} estimates from each \ac{sa}. For a given \ac{sa} with $U$ sources, the loss is defined as

\begin{equation}
    \mathcal{L}_{n} = \min_{p \in \mathcal{P}} \frac{1}{U} \sum_{u=1}^{U} 
    \left[ \operatorname{mod}\Bigl( \hat{\varphi}_{n,p(u)} - \varphi_u, \pi \Bigr) \right]^2,
\end{equation}
where $\varphi_k$ are the ground-truth angles, $\hat{\varphi}_{n,p(u)}$ are the predicted angles, $\mathcal{P}$ is the set of all permutations of $U$ elements, and $\operatorname{mod}(\cdot, \pi)$ ensures the angular difference is wrapped into $[-\pi, \pi]$ to account for the periodicity of angles. The total training loss is then computed as the average across all \acp{sa},
\begin{equation}
    \mathcal{L} = \frac{1}{N} \sum_{n=1}^N \mathcal{L}_{n}.
\end{equation}
This \ac{sa}-wise processing strategy enhances generalization through weight sharing and enables the model to focus on local spatial statistics.

\section{Numerical Results}
\label{results}
The proposed model was implemented in \texttt{PyTorch} and operates at a frequency of 142~GHz. In all experiments, there are $N=15$ \acp{sa}, each containing $M = 25$ antennas and observing $T = 10$ time samples, with the number of sources set to $U=2$. Autocorrelation matrices were computed at 12 distinct time lags. The inter-\ac{sa} spacing is $\Delta = M\lambda/2$, and the inter-element spacing is $\delta = \lambda/2$. The feature space and the transformer feed-forward layers each have a dimensionality of 128, and the encoder comprises three transformer layers. The model was trained for 25 epochs using the AdamW optimizer with an initial learning rate of $5 \times 10^{-4}$, a weight decay of $10^{-3}$, and gradient norm clipping set to 10. The learning rate was decayed following a cosine annealing schedule, reaching a minimum value of $5 \times 10^{-5}$ by the final epoch.
\begin{figure}[t]
    \centering
    \includegraphics[width=0.95\linewidth]{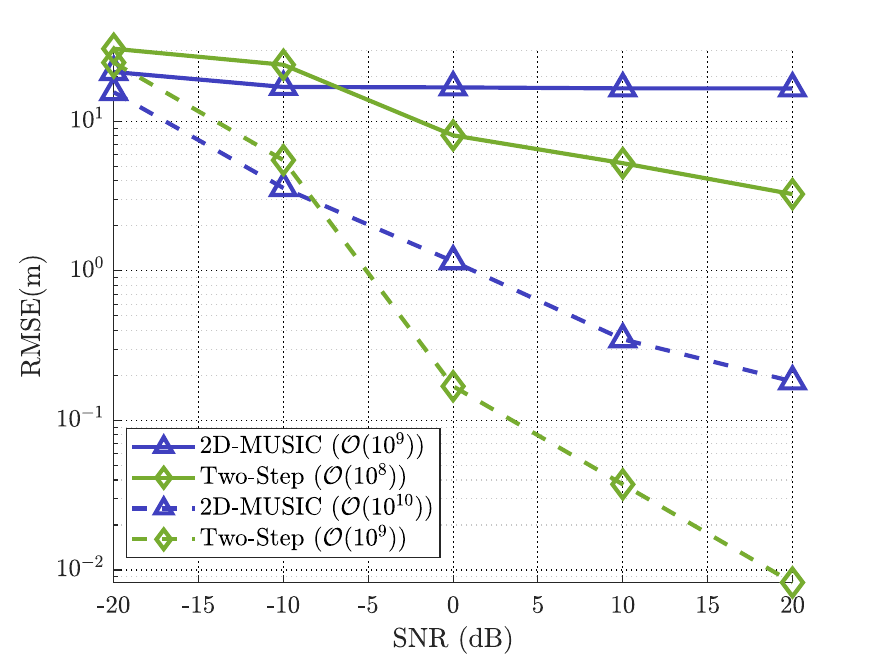}
    \caption{Comparison of localization accuracy and computational complexity between the proposed framework and 2D-\ac{music}.}
    \label{fig:2D_1D_music}
\end{figure}

Training was performed on datasets with a mix of perfectly coherent and non-coherent sources with \acp{snr} of $10$~dB and $-10$~dB. The model was first trained using data from the center \ac{sa}, and the resulting weights were subsequently used to initialize training across \ac{sa}. 

The angle estimation error is evaluated using the \ac{prmse}, defined as \(\text{PRMSE} = \min_{p \in \mathcal{P}} \sqrt{\frac{1}{U} \sum_{u=1}^{U} ( \hat{\varphi}_{p(u)} - \varphi_u )^2}\), where angular differences are wrapped to \([-\pi,\pi]\). Range error is defined similarly without permutation. This metric accounts for the arbitrary labeling of sources.

For complexity analysis of Algorithm~\ref{alg:multiuser_two_step}, using the parameters \(N = 15\), \(M = 25\), \(U = 10\), \(\varphi \in [-\pi/3, \pi/3]\), \(\Delta \varphi = 0.001 \implies N_\varphi \approx 2094\), \(r \in [2, 80]~\text{m}\), \(\Delta r = 0.1~\text{m} \implies N_r = 780\), and \(T = 100\) snapshots, the full 2D-MUSIC requires \(\sim 2.30 \cdot 10^{11}\) operations, whereas the proposed hierarchical MUSIC-MUSIC reduces complexity to \(\sim 2.32 \cdot 10^{7}\) operations (approximately four orders of magnitude improvement). Hierarchical Root-MUSIC-MUSIC further reduces complexity to \(\sim 3.57 \cdot 10^{6}\) operations.

\begin{figure*}[ht]
    \centering
    \subfloat[AoA estimation performances.]{
        \label{fig:coherent_case_rmse_radians}
        \includegraphics[width=0.45\linewidth]{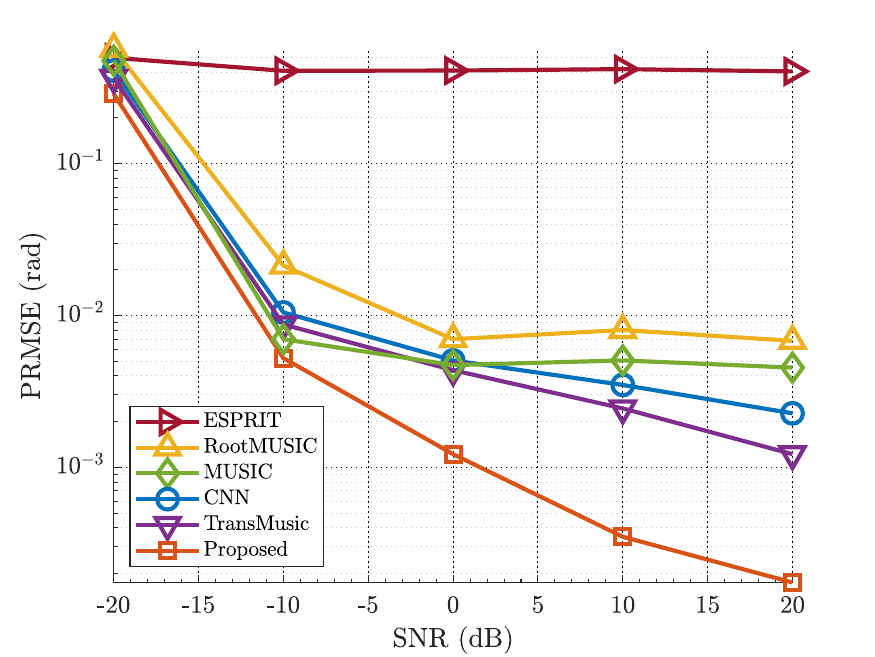}
    }\hfill
    \subfloat[Range estimation performance.]{
        \label{fig:range_coherent}
        \includegraphics[width=0.45\linewidth]{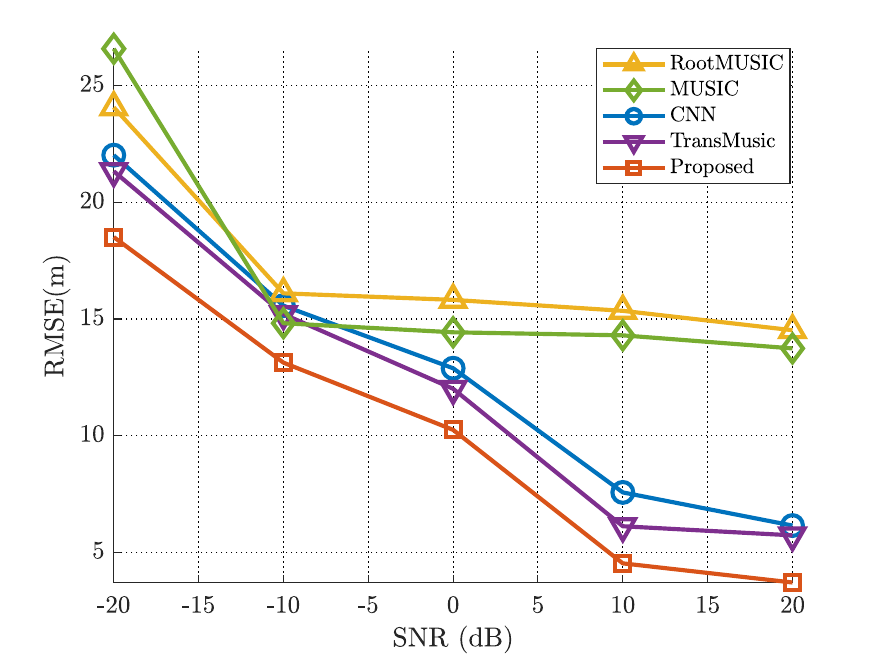}
    }
    \caption{Localization under coherent source conditions for different methods.}
    \label{fig:coherent_combined}
    \vspace{-2mm}
\end{figure*}

Fig.~\ref{fig:2D_1D_music} compares the location estimation performance of the proposed hierarchical algorithm~\ref{alg:multiuser_two_step} using 1D \ac{music} with traditional 2D‑\ac{music}. The proposed method achieves a tenfold reduction in computational complexity while significantly improving localization performance at mid‑to‑high \ac{snr} levels. At very low \ac{snr}, a minor performance gap is observed, attributable to error propagation from Step 1 to Step 2. At moderate‑to‑high \ac{snr}, accurate Step‑1 angles enable full beamforming gain, making Step‑2 range estimation robust.

To assess the effectiveness of the transformer model, we test it under a coherent sources scenario, as illustrated in Fig.~\ref{fig:coherent_combined}. The proposed transformer‑based method is compared against classical subspace and machine learning algorithms that were applied using algorithm~\ref{alg:multiuser_two_step}. In Fig.~\ref{fig:coherent_case_rmse_radians}, which shows the angular estimation performance, the proposed method (with covariance correction) achieves the lowest \ac{prmse} across all \ac{snr} values, significantly outperforming other approaches. At an \ac{snr} of $5$ dB, the \ac{rmse} drops from $0.005$ for MUSIC and approximately $0.008$ for Root‑MUSIC to $0.0007$ with the proposed approach, yielding an $85\%$ improvement over MUSIC. In Fig.~\ref{fig:range_coherent}, which depicts range estimation \ac{rmse}, the proposed method reduces the range \ac{rmse} by about $6.5~\mathrm{m}$ relative to MUSIC at $5$ dB \ac{snr}, demonstrating clear superiority over traditional methods.

\section{Conclusion}
\label{conclusion}
We proposed a transformer-aided hierarchical framework with \ac{sa} processing that reduces 2D angle-range search to sequential 1D searches, achieving a $10^{4}$ complexity reduction over full 2D MUSIC. A supervised covariance correction model enhances robustness under coherent sources, enabling scalable THz \ac{nf} localization for ultra-massive arrays and bridging model-based with data-driven approaches. The proposed method improves angular \ac{prmse} by $85\%$ and reduces range \ac{rmse} by $6.5~\mathrm{m}$ at $5~\mathrm{dB}$ \ac{snr}.

\bibliographystyle{IEEEtran}
\bibliography{IEEEabrv,ref}

\end{document}